\documentclass[conference,a4paper]{IEEEtran} 
\usepackage[utf8]{inputenc} 
\usepackage[T1]{fontenc}
\usepackage{url} 
\usepackage{ifthen}
\usepackage{cite} 
\usepackage[cmex10]{amsmath} 
\newtheorem{theorem}{Theorem}
\newtheorem{lemma}{Lemma}
\newtheorem{definition}{Definition}

\usepackage{amssymb}

\makeatletter
\def\@begintheorem#1#2{\@IEEEtmpitemindent\itemindent\relax\topsep 0pt\rmfamily\trivlist%
    \item[]{\indent{\bfseries #1\ #2}:} \itemindent\@IEEEtmpitemindent\relax\itshape}
\def\@opargbegintheorem#1#2#3{\@IEEEtmpitemindent\itemindent\relax\topsep 0pt\rmfamily \trivlist%
    \item[]{\indent {\bfseries #1\ #2}\ (#3):} \itemindent\@IEEEtmpitemindent\relax\itshape}
\def\@endtheorem{\endtrivlist}
\makeatother
\newtheorem{remark}{Remark}
\let\oldremark\remark
\def\remark{\oldremark\upshape}

\newcommand{\E}{\mathbb{E}}

\newcommand{\R}{\mathbb{R}}
\newcommand{\C}{\mathbb{C}}
\let\tilde\widetilde
\let\hat\widehat

\newcommand{\A}{\mathbf{A}}
\newcommand{\B}{\mathbf{B}}
\newcommand{\D}{\mathbf{D}}
\newcommand{\U}{\mathbf{U}}
\renewcommand{\L}{\mathbf{L}}
\newcommand{\I}{\mathbf{I}}
\newcommand{\K}{\mathbf{K}}
\newcommand{\M}{\mathbf{M}}
\newcommand{\W}{\mathbf{W}}
\newcommand{\w}{\mathbf{w}}

\newcommand{\tr}{\mathrm{tr}}
\newcommand{\diag}{\mathrm{diag}}

\renewcommand{\Re}{\operatorname{Re}}
\renewcommand{\Im}{\operatorname{Im}}

\makeatletter 
\renewcommand*\env@matrix[1][*\c@MaxMatrixCols c]{%
  \hskip -\arraycolsep
  \let\@ifnextchar\new@ifnextchar
  \array{#1}}
\makeatother
\newenvironment{Smallmatrix}[1]
  {\arraycolsep=3pt\footnotesize\array{#1}}
  {\endarray}

\begin{document}
\title{Equality in the Matrix Entropy-Power Inequality and Blind Separation of Real and Complex sources} 

\author{%
  \IEEEauthorblockN{Olivier Rioul}
  \IEEEauthorblockA{LTCI, T\'el\'ecom Paris\\
                    Institut Polytechnique de Paris,
                    75013, Paris, France\\
                    Email: olivier.rioul@telecom-paristech.fr}
  \and
  \IEEEauthorblockN{Ram Zamir}
  \IEEEauthorblockA{EE - Systems Department\\
                    Tel Aviv University, 
                    Tel Aviv, Israel\\
                    Email: zamir@eng.tau.ac.il}
}

\maketitle

\begin{abstract}
The matrix version of the entropy-power inequality for real or complex coefficients and variables is proved using a transportation argument that easily settles the equality case. An application to blind source extraction is given.
\end{abstract}


\section{Introduction} 

Consider random variables with densities that are continuous 
and positive inside their support interval, with zero mean and finite differential entropies.
%
The entropy power inequality (EPI) was stated by 
Shannon~\cite{Shannon48} in 1948 and is well known to be equivalent to the 
following minimum entropy inequality~\cite{Lieb78,DemboCoverThomas91,Rioul11}:
\begin{equation}\label{epi1}
h(a_1X_1+a_2X_2) \geq h(a_1X_1^*+a_2X_2^*) 
\end{equation}
for any real numbers $a_1,a_2$ and any independent real random variables $X_1,X_2$, 
where $X^*_1,X^*_2$ are independent normal random variables having the same 
entropies as $X_1,X_2$:
\begin{equation}\label{epise}
h(X^*_1)=h(X_1) \qquad h(X^*_2)=h(X_2).
\end{equation}
Equality holds in~\eqref{epi1} if and only if either $a_1a_2=0$ or $X_1,X_2$ are 
normal. 
Recently, a normal transport argument was used in~\cite{Rioul17} to provide a simple 
proof of Shannon's EPI, including the necessary and sufficient condition for equality. 

Shannon's EPI was generalized to a matrix version~\cite{ZamirFeder93,ZamirFeder93b}:
\begin{equation}\label{mepi}
h(\A X) \geq h(\A X^*) 
\end{equation}
for any $m\times n$ matrix $\A$ and any random (column) vector 
$X=(X_1,X_2,\ldots,X_n)^t$
of independent components $X_i$, where 
$X^*=(X^*_1,X^*_2,\ldots,X^*_n)^t$
is a normal vector with independent components $X^*_i$ of the same entropies:
\begin{equation}\label{sameentropies}
h(X^*_i)=h(X_i) \qquad (i=1,\ldots,n).
\end{equation}
Available proofs of~\eqref{mepi} are either by double induction on $(m,n)$~\cite{ZamirFeder93} or by integration over a path of Gaussian perturbation of the corresponding inequality for Fisher’s information using de Bruijn's identity~\cite{ZamirFeder93b} or via the I-MMSE relation~\cite{GuoShamaiVerdu06}.
A necessary and sufficient condition for equality in~\eqref{mepi} has not been settled so far, however, by the previous methods. Such a condition is important in applications such as blind source separation (BSS) based on minimum entropy~\cite{Vrins07}. 
Also, BSS may involve real or complex signals~\cite{Cardoso93} and  minimum entropy methods for complex sources would require the extension of EPIs to complex-valued variables and coefficients.

In this paper, we adapt the proof of~\cite{Rioul17} to the matrix case and derive~\eqref{mepi} with a normal transport argument. This allows us to easily settle the equality case: We define the notion of ``recoverability'' and show that equality holds in~\eqref{mepi} if all unrecoverable components of $X$ present in $\A X$ are normal. We then extend the proofs to complex-valued~$\A$ and~$X$. As an application, we derive the appropriate contrast functions for partial BSS (a.k.a. blind source extraction) where $m$ out of $n$ independent sources are to be extracted.

\section{A simple proof of the matrix EPI by transport}\label{proofsection}

We extend the proof in~\cite{Rioul17} to the matrix EPI, based on the same ingredients: (a)~a transportation argument from normal variables, that takes the form of a simple change of variables; (b)~a rotation performed on i.i.d. normal variables, which preserves the i.i.d. property; (c)~concavity of the logarithm, appropriately generalized to the matrix case. The proof breaks into several elementary steps:

\subsection{Reduce to full rank $m<n$}\label{m<n}\label{A}

If the rank of $\A$ is $<m$ then some rows are linearly dependent, there is a deterministic relation between some components of $\A X$ and $\A X^*$ and equality  $h(\A X) = h(\A X^*) =-\infty$ holds trivially. 
Thus we can assume that $\A$ is of full rank $m\leq n$.  If $\A$ has rank $m=n$ then $\A$ is invertible and by the change of variable formula in the entropy~\cite[\S~20.9]{Shannon48},
$h(\A X) =h(X) + \log|\A| = h(X^*)+\log|\A|= h(\A X^*)$ where $|\A|$ denotes the absolute value of the determinant of~$\A$. Therefore, one may always assume that  $\A$ has full rank $m\!<\!n$.

\subsection{Reduce to equal individual entropies} \label{h=h}\label{B}

Without loss of generality, one may assume that the components of $X$ have equal 
entropies. For if it were not the case, then by the scaling property of 
entropy\cite[\S~20.9]{Shannon48}, one can find non zero coefficients $
\delta_j$ (e.g., $\delta_j=\exp 
h(X_j)$) such that all  
$X'_j=X_j/\delta_j$ have equal 
entropies. 
Then applying~\eqref{mepi} to $X'=(X'_1,\ldots,X'_n)^t$ and matrix $\A \pmb{\Delta}$ 
where $\pmb{\Delta}$ is a diagonal matrix with diagonal elements $\delta_j$,
gives the desired EPI.

Notice that with the additional constraint that the $X_j$ have equal entropies, we have $h(X^*_1)= h(X^*_2)=\cdots=h(X^*_n)=h(X_1) =h(X_2)=\cdots=h(X_n)$: The independent zero-mean normal variables $X^*_j$ also have equal entropies, and are, therefore, independent and identically distributed (i.i.d.).

\subsection{Reduce to orthonormal rows} \label{C}

Without loss of generality, one may assume that the rows of $\A$ are orthonormal.
For if it were not the case, one can orthonormalize the rows by a Gram-Schmidt  process. 
This amounts to multiplying $\A$ on the left by an lower-triangular invertible matrix 
$\L$.
Thus, one can apply~\eqref{mepi} for matrix $\A'=\L\A$. Again by the change of 
variable in the entropy\cite[\S~20.9]{Shannon48},
$h(\A' X)= h(\A X) + \log|\L|$ and $h(\A' X^*)=  h(\A X^*)+\log|\L|$. The terms $\log|
\L|$ cancel to give the desired EPI. Thus we are led to prove~\eqref{mepi}
for an $m\times n$ matrix $\A$ with orthonormal rows ($\A\A^t=\I_m$, the $m\times m$ identity matrix).

\subsection{Complete the orthogonal matrix}\label{D}

Extend $\A$ by adding $n-m$ orthonormal rows of a complementary matrix $\A'$ such 
that 
$\bigl(\!
\begin{smallmatrix}
\A\hphantom{'}\\[0.5ex]\hline \A' 
\end{smallmatrix}
\!\bigr)$
is an $n\times n$ orthogonal matrix, and define the Gaussian vector $\Bigl(\!
\begin{smallmatrix}
\tilde{X}\hphantom{'}\\[0.5ex]\hline \vphantom{^{2^2}}\tilde{X}' 
\end{smallmatrix}
\!\Bigr)$ as
\begin{equation}
\begin{pmatrix}
\tilde{X}\\\hline \vphantom{1^{1^{1^1}}}\tilde{X}' 
\end{pmatrix}=
\begin{pmatrix}
\A\hphantom{'}\\\hline \A' 
\end{pmatrix}
X^*.
\end{equation}
Since the components of $X^*$ are i.i.d. normal and $\bigl(\!
\begin{smallmatrix}
\A\hphantom{'}\\[0.5ex]\hline \A' 
\end{smallmatrix}
\!\bigr)$ is orthogonal, the components of $\Bigl(\!
\begin{smallmatrix}
\tilde{X}\hphantom{'}\\[0.5ex]\hline \vphantom{^{2^2}}\tilde{X}' 
\end{smallmatrix}
\!\Bigr)$
are also i.i.d. normal. In particular the subvectors $\tilde{X}$ and $\tilde{X}'$ are independent.
The inverse transformation is the transpose:
\begin{equation}\label{invortho}
X^* =  \begin{pmatrix}
\A^t \;\Big|\; {\A'\,}^t 
\end{pmatrix}
\begin{pmatrix}
\tilde{X}\\\hline  \vphantom{1^{1^{1^1}}}\tilde{X}' 
\end{pmatrix}=\A^t  \tilde{X} + {\A'\,}^t  \tilde{X}' .
\end{equation}

\subsection{Apply the normal transportation}\label{E}

\begin{lemma}[Normal Transportation~\cite{Rioul17,Rioul17a}]\label{nt}
Let $X^*\in\R$ be a scalar normal random variable. For any continuous density~$f$,  there exists a differentiable 
transformation $T\colon\R\!\to\!\R$ with positive derivative $T'\!>\!0$ such that 
$X=T(X^*)$ has density~$f$. 
\end{lemma}

From Lemma~\ref{nt}, we can assume that the components of $X=(X_1,X_2,\ldots,X_n)^t$ and $X^*=(X^*_1,X^*_2,\ldots,X^*_n)^t$ are such that
$X_j=T_j(X^*_j)$ for all $j=1,2,\ldots,n$, 
where the $T_j$'s are transformations with positive derivatives $T'_j>0$. For ease of 
notation define
\begin{equation}
T(X^*)=\bigl(T_1(X^*_1), T_2(X^*_2),\ldots,T_n(X^*_n)\bigr)^t
\end{equation}
Thus $T\colon\R^n\to \R^n$ is a transformation whose Jacobian matrix is diagonal with 
positive diagonal elements:
\begin{equation}\label{jacobiandiag}
T'(X^*) =  \diag \bigl(T'_1(X^*_1), \ldots, T'_n(X^*_n)\bigr)
.
\end{equation}
Now~\eqref{mepi} can be written in terms of the normal variables only:
\begin{equation}
h\bigl(\A T(X^*)\bigr) \geq h(\A X^*) 
\end{equation}
and by~\eqref{invortho} it can also be written in term of the tilde normal variables:
\begin{equation}\label{mepitilde}
h\bigl(\A T(\A^t  \tilde{X} + {\A'\,}^t  \tilde{X}' )\bigr) \geq h(\tilde{X}).
\end{equation}

\subsection{Conditioning on the complementary variables}\label{F}

Since conditioning reduces entropy~\cite[\S~20.4]{Shannon48},
\begin{equation}\label{crevect}
h\bigl(\A T(\A^t  \tilde{X} \!+\! {\A'\,}^t  \tilde{X}' )\bigr) 
\geq
h\bigl(\A T(\A^t  \tilde{X} \!+\! {\A'\,}^t  \tilde{X}' ) \mid  \tilde{X}' \bigr).
\end{equation}

\subsection{Make the change of variable}\label{G}

By the change of variable formula in the entropy~\cite[\S~20.8]{Shannon48}, 
$h(X_j)=h(T_j(X^*_j)) =h(X^*_j)+ \E \log T_j'(X^*_j)$ and, therefore, by~\eqref{sameentropies},
\begin{equation}\label{ElogT'zero}
\E \log T_j'(X^*_j)=0\qquad (j=1,2,\ldots,n). 
\end{equation}

By the change of variable formula (vector case)~\cite[\S~20.8]{Shannon48} in the 
conditional entropy in the r.h.s. of~\eqref{crevect},
\begin{align}
h\bigl(\A &T(\A^t  \tilde{X} + {\A'\,}^t  \tilde{X}' ) \mid  \tilde{X}' \bigr)\notag\\
&= h( \tilde{X}\mid  \tilde{X}' ) +\E \log | \A T'(\A^t  \tilde{X} + {\A'\,}^t  \tilde{X}' ) 
\A^t|\\
&= h(\tilde{X}) + \E \log | \A T'(X^*) \A^t|
\label{chainrulejacobian}
\end{align}
where we have used that $\tilde{X}$ and $\tilde{X}'$ are independent.

\subsection{Apply the concavity of the logarithm}\label{H}

The following lemma was stated in~\cite{ZamirFeder93b} as a 
consequence of~\eqref{mepi}. A direct proof was given in~\cite{GuoShamaiVerdu06}, and is simplified here.

\begin{lemma}\label{logconcmatrixlemma}
For any $m\times n$ matrix $\A$ with orthonormal rows and any diagonal matrix $
\pmb{\Lambda}=\diag(\lambda_1,\ldots,\lambda_n)$ with positive diagonal 
elements $\lambda_j>0$,
\begin{equation}\label{logconcmatrix}
\log|\A \pmb{\Lambda} \A^t | \geq \tr (\A [\log \pmb{\Lambda}] \A^t)
\end{equation} 
where $\log \pmb{\Lambda}=\diag(\log\lambda_1,\ldots,\log\lambda_n)$ and $\tr(\cdot)$ denotes the trace.
\end{lemma}
Equality holds e.g. when the $\lambda_j$'s are equal. The precise equality case will appear elsewhere.
\begin{IEEEproof}
It is easily checked that $\A \pmb{\Lambda} \A^t$ is positive definite and that
both sides of~\eqref{logconcmatrix} do not change if we replace $\A$ by $\U\A$ where $\U$ is any $m\times m$ orthogonal matrix.
%
Choose $\U$ as an orthogonal eigenvector matrix of $\A 
\pmb{\Lambda} \A^t$, so that $\U\A \pmb{\Lambda} \A^t\U^t$ is diagonal with 
positive diagonal elements and $\U\A$ still has orthonormal rows.

Thus, substituting $\U\A$ for $\A$ we may always assume that $\A \pmb{\Lambda} \A^t$ is diagonal with diagonal entries equal to $\sum_{j=1}^n A^2_{ij}\lambda_j $ for $i=1,2,\ldots,m$, where $A_{i,j}$ denotes the entries of $\A$. Then
\begin{align}
\log|\A \pmb{\Lambda} \A^t | 
&= \sum_{i=1}^m \log \sum_{j=1}^n A^2_{ij}\lambda_j\\
&\geq \sum_{i=1}^m\sum_{j=1}^n A^2_{ij} \log\lambda_j \label{logconcUA}\\
&=\tr (\A [\log \pmb{\Lambda}] \A^t).
\end{align}
where~\eqref{logconcUA} follows from Jensen's inequality and the concavity of the logarithm,   
since $\A$ has orthonormal rows.
\end{IEEEproof}
From Lemma~\ref{logconcmatrixlemma} and~\eqref{ElogT'zero} we obtain
\begin{align}\label{dettraceineq}
 \E \log | \A T'(X^*) \A^t| &\geq \E\;\tr ( \A [\log T'(X^*)] \A^t) 
 \\&= \tr ( \A \E[\log 
T'(X^*)] \A^t) = 0.
\end{align}
Combining this with~\eqref{crevect}--\eqref{chainrulejacobian} proves~\eqref{mepitilde} and the desired matrix EPI~\eqref{mepi}.

\section{The Equality Case}\label{eqsection}
To settle the equality case in~\eqref{mepi}, from the remarks in \S~\ref{m<n} we may already assume that $\A$ has full rank $m<n$.
\begin{definition}\label{presrecov}
A component $X_j$ of $X$ is%
\vspace*{-1ex}
\begin{itemize}
\item \emph{present} in $\A X$ if $\A X$ depends on $X_j$;
\item \emph{recoverable} from $\A X$ if there exists a row vector $b$ such that
$b\!\cdot\! (\A X)  = X_j$. 
\end{itemize}
\end{definition}
\begin{remark}
Since the considered variables are not deterministic, Definition~\ref{presrecov} depends only on the matrix~$\A$: $X_j$ is present in $\A X$ if and only if the $j$th column of $\A$ is not zero; and $X_j$ is recoverable from $\A X$ if and only if there exists $b$ such that $b\A=(0,\ldots,0,1,0,\ldots,0)$ with $1$ in the $j$th position. A recoverable component is necessarily present. 
\end{remark}
\begin{remark}\label{deletezero}
Without loss of generality we always omit the components that are not present in $\A X$ and their associated zero columns of $\A$ without affecting the entropy $h(\A X)$. 
\end{remark}
\begin{remark}
Definition~\ref{presrecov} is also invariant by left multiplication of $\A$ by any $m\times m$ invertible matrix $\B$: if the $j$th column of $\A$ is zero, so is the $j$th column of $\B\A$; and $b\A=(0,\ldots,0,1,0,\ldots,0)$ implies $(b\B^{-1}) (\B \A) = (0,\ldots,0,1,0,\ldots,0)$.
\end{remark}

The following property was used in~\cite[Appendix]{ZamirFeder98} for deriving a sufficient condition for equality in a matrix form of the Brunn–Minkowski inequality,
which is the analog of the EPI for R\'enyi entropies of order zero~\cite{DemboCoverThomas91}.
\begin{lemma}\label{lem-can}
Reordering the components of $X$ if necessary so that the first $r$ components are recoverable and the last $n-r$ components are unrecoverable, we may always put $\A$ in the canonical form
\begin{equation}\label{canonical}
\A = \begin{pmatrix}[c|c]
\I_r & \pmb{0} \\\hline
\pmb{0} & \;\A_u\;
\end{pmatrix}
\end{equation}
where $\A_u$ is an $(m-r)\times(n-r)$ matrix.
The number $r$ of recoverable components is the \emph{maximum} number such that $\A$ can be put in the form~\eqref{canonical} by left multiplication by an invertible matrix.
\end{lemma}
\begin{IEEEproof}
Write $X=(X_r\mid X_u)^t$ where $X_r$ has recoverable components and $X_u$ has unrecoverable ones. 
By Definition~\ref{presrecov} (recoverability) there exists a $r\times m$ matrix $\B_r$ such that $\B_r\A=(\I_r\mid 0)$. Since $\B_r$ must have rank $r$, this shows in particular that $r\leq m$: no more than $m$ components can be recovered from the $m$ linear mixtures. We can use $m-r$ additional row operations so that 
$
\bigl(
\begin{smallmatrix}
\B_r\\[0.5ex]\hline \B_u
\end{smallmatrix}
\bigr)
\A
=
\bigl(
\begin{Smallmatrix}{c|c}
\I_r & \pmb{0} \\\hline
\pmb{0} & \A_u
\end{Smallmatrix}
\bigr)
$
is of the desired form. Since $\B=\bigl(
\begin{smallmatrix}
\B_r\hphantom{'}\\[0.5ex]\hline \B_u 
\end{smallmatrix}
\bigr)$ is an $m\times m$  invertible matrix, 
by the change of variable formula in the entropy~\cite[\S20.9]{Shannon48},
$h(\B\A X) =h(\A X) + \log|\B|$ and $h(\B\A X^*)=h(\A X^*)+\log|\B|$.
Therefore, the matrix EPI~\eqref{mepi} is equivalent to the one obtained by substituting
$\B\A=
\bigl(
\begin{Smallmatrix}{c|c}
\I_r & \pmb{0} \\\hline
\pmb{0} & \A_u
\end{Smallmatrix}
\bigr)
$
 for~$\A$. Clearly, $r$ is maximum in this expression since otherwise one could recover more than $r$ components, hence transfer
some of the components from the $\A_u$ block to the $\I_r$ block.
\end{IEEEproof}
We can now settle the equality case in~\eqref{mepi}.
\begin{theorem}\label{thm-eq}
Equality holds in~\eqref{mepi} if and only if all unrecoverable components present in $\A X$ are normal.  
\end{theorem}
\begin{IEEEproof}
Write $X=(X_r\mid X_u)^t$ as in the proof of Lemma~\ref{lem-can} and accordingly write $X^*=(X^*_r\mid X^*_u)^t$.
If $\A$ is in canonical form~\eqref{canonical}, then~\eqref{mepi} reads
\begin{equation}
h(X_r)+h(\A_u X_u) \geq h(X^*_r) + h(\A_u X^*_u).
\end{equation}
where 
$h(X_r)=\sum_{j=1}^r h(X_j) =\sum_{j=1}^r h(X^*_j) = h(X^*_r)$.
The announced condition is, therefore, sufficient: if $X_u$ is normal with (zero-mean) components satisfying~\eqref{sameentropies}, then $X_u$ is identically distributed as $X_u^*$ and $h(\A_u X_u)=h(\A_u X^*_u)$.

Conversely, suppose that~\eqref{mepi} is an equality with $\A$ as in~\eqref{canonical}. From \S~\ref{proofsection}~C, we may assume (applying row operations of a Gram-Schmidt process if necessary) that $\A$ has orthonormal rows in~\eqref{canonical}, that is, $\A_u\A_u^t=\I_{m-r}$. Then equality holds in~\eqref{mepi} if and only if both~\eqref{crevect} and~\eqref{dettraceineq} are equalities.

Consider equality in~\eqref{dettraceineq} which results from the application of Lemma~\ref{logconcmatrixlemma} (inequality~\eqref{logconcmatrix}) to $\pmb{\Lambda}=T'(X^*)$. We have
\begin{equation}
\A \pmb{\Lambda} \A^t = \begin{pmatrix}[c|c]
\pmb{\Lambda}_r & \pmb{0} \\\hline
\pmb{0} & \;\A_u\pmb{\Lambda}_u\A_u^t\;
\end{pmatrix} 
\end{equation}
where $\pmb{\Lambda}_r=\diag(\lambda_1,\ldots,\lambda_r)$ and $\pmb{\Lambda}_u=\diag(\lambda_{r+1},\ldots,\lambda_n)$.
Thus, we may choose $\U$ in the proof of Lemma~\ref{logconcmatrixlemma} in the form $\U=
\bigl(
\begin{Smallmatrix}{c|c}
\I_r & \pmb{0} \\\hline
\pmb{0} & \U_u
\end{Smallmatrix}
\bigr)
$ where $\U_u$ is an $(m-r)\times(m-r)$ orthogonal matrix such that $\U_u \A_u\pmb{\Lambda}_u\A_u^t\U_u^t$ is diagonal. Then 
$
\U\A=\bigl(
\begin{Smallmatrix}{c|c}
\I_r & \pmb{0} \\\hline
\pmb{0} & \U_u\A_u
\end{Smallmatrix}
\bigr)
$
 is still of the form~\eqref{canonical} where $\U_u\A_u$ has orthonormal rows. 
 
 Therefore, equality in~\eqref{logconcmatrix} is equivalent to equality in~\eqref{logconcUA} where we may again assume that $\A$ is of the form~\eqref{canonical} where $r$ is maximal and $\A_u$ has orthonormal rows.
By Remark~\ref{deletezero}, we may assume that all columns of $\A_u$ are nonzero.
Notice that any row of $\A_u$ in~\eqref{canonical} should have \emph{at least two nonzero elements}. Otherwise, there would be one row of $\A_u$ of the form $(0,\ldots,0,\pm 1,0,\ldots,0)$ with the nonzero element in the $j$th position. Since the rows are orthonormal, the other elements in the $j$th column would necessarily equal zero, and the corresponding component of $X$ would be recoverable, which contradicts the maximality of $r$.

Now since the logarithm is strictly concave, equality holds in~\eqref{logconcUA} if and only if for 
all $i=1,2,\ldots,m$,  all the $\lambda_j$ for which ${A}_{i,j}\ne 0$ are equal.
Because no column of $\A_u$  is zero and any row of $\A_u$ in~\eqref{canonical} has at least two nonzero elements, this implies that for any $j$ such that $r<j\leq n$, $\lambda_j$  is equal to another $\lambda_k$ where $r<k\leq n$, $k\ne j$.
Since Lemma~\ref{logconcmatrixlemma} was applied to $\pmb{\Lambda}=T'(X^*)$ it follows that 
\begin{equation}\label{dereqae}
T'_j(X^*_j)=T'_k(X^*_k) \text{ a.e.} \quad (r<j,k\leq n)
\end{equation}
Because $X^*_j$ and $X^*_k$ are independent, this implies that both $T'_j(X^*_j)$ 
and $T'_k(X^*_k)$ are constant and equal a.e., hence $T'_j=T'_k=c$ for some 
constant\footnote{This is similar to what appeared in an earlier transportation proof of the EPI~\cite{Rioul17}. By~\eqref{ElogT'zero}, we necessarily have $c=1$ if we assume that all individual entropies are equal as in \S~\ref{h=h}. } $c$. Therefore $T_j$ is linear and $X_j=T_j(X^*_j)$ is normal for all $r<j\leq n$. 
This completes the proof.\footnote{%
This implies, in particular, that equality in~\eqref{dettraceineq} implies equality in~\eqref{crevect}. This can also be seen directly: if $T'_j=1$ for all $r<j\leq n$, then for $\A$ of the form~\eqref{canonical} in~\eqref{crevect}, 
$\A_u T(\A_u^t  \tilde{X} \!+\! {\A'\,}^t  \tilde{X}' ) = \tilde{X}$ is independent of $\tilde{X}' $.}
\end{IEEEproof}

\section{Extension to Complex Matrix and Variables}

A complex random variable $X\in\C$ can always be viewed as a two-dimensional real random vector $\hat{X}=\bigl(\!
\begin{smallmatrix}
\Re X \\ \Im X 
\end{smallmatrix}\!\bigr)\in\R^2$.
Therefore, by the vector form of the EPI~\cite{Lieb78,DemboCoverThomas91,Rioul11}, \eqref{epi1} holds for scalar coefficients $a_1,a_2\in\R$ when $X_1,X_2\in\C$ are independent complex random vectors and $X^*_1,X^*_2\in\C$ are independent white normal random vectors  satisfying~\eqref{epise}. Here ``white  normal'' $X^*\in\C$ amounts to say that $X^*$ is \emph{proper} normal or \emph{circularly symmetric} normal~\cite{Picinbono94}  (\emph{c-normal} in short): $X^*\sim\mathcal{CN}(0,\sigma^2)$, that is, $\hat{X^*} \sim\mathcal{N}(0,\sigma^2\I_2)$.

That \eqref{epi1} also holds for \emph{complex} coefficients $a_1,a_2\in\C$ is less known but straightforward. To see this, define\footnote{There is an ambiguity of notation easily resolved from the context: $\hat{a}$ is a matrix when $a$ is a constant and $\hat{X}$ is a vector when $X$ is random.}
$\hat{a}=\bigl(\!
\begin{smallmatrix}
\Re a & -\Im a \\ \Im a & \hphantom{-}\Re b
\end{smallmatrix}\!\bigr)$
for any $a\in\C$, 
so that $\hat{a X} = \hat{a} \hat{X}$. Then $h(aX)=h(\hat{a}\hat{X})=h(\hat{X})+\log|\hat{a}| = h(X)+\log|a|^2$.
Hence~\eqref{epise} implies $h(a_1X_1)=h(a_1X^*_1)$ and $h(a_2X_2)=h(a_2X^*_2)$. 
In addition, if $X^*\!\sim\!\mathcal{CN}(0,\sigma^2)$ then $aX^*\!\sim\!\mathcal{CN}(0,|a|^2\sigma^2)$. Therefore, by the vector EPI applied to $a_1X_1$ and $a_2X_1$ we see that \eqref{epi1} holds for complex coefficients $a_1,a_2\in\C$ when $X^*_1,X^*_2$ are independent c-normal variables  satisfying~\eqref{epise}.  


The extension of the matrix EPI~\eqref{mepi} to \emph{complex} $\A$ and $X$ is more involved.
We need the following notions~(see, e.g., \cite{ErikssonKoivunen06} and~\cite[chap. 10]{Rioul08book}). 
Define $\hat{X}\in\R^{2n}$ by stacking the $\hat{X_i}$ for each component $X_i\in\C$ of $X\in\C^n$, and 
define $\hat{\A}$ as the $2m \times 2n$ real matrix with $2\times 2$ entries $\hat{A_{i,j}}$ where $A_{i,j}$ are the complex entries of $\A$. It is easily checked that $\hat{\A X}=\hat{\A} \hat{X}$, $\hat{\A\B}=\hat{\A}\hat{\B}$, $\hat{\A^\dag}=\hat{\A}^t$ where $\A^\dag$ is the conjugate transpose, and $|\hat{\A}|=|\A|^2$ where $|\A|$ denotes the modulus of the determinant of $\A$.

We also need the following extension of Lemma~\ref{nt}:
\begin{lemma}[2D Brenier Map\cite{Brenier91,McCann95}]\label{bm}
Let $\hat{X^*}\in\R^2$ be a (white) normal random vector. For any given continuous density~$f$ over $\R^2$,  there exists a differentiable 
transformation $T\colon\R^2\!\to\!\R^2$ with symmetric positive definite Jacobian $T'$ (noted $T'\!>\!0$) such that  $\hat{X}=T(\hat{X^*})$ has density~$f$. 
\end{lemma}
Courtade et al.~\cite{CourtadeFathiPananjady18} noted that the Brenier map can be used in the transportation proof of~\cite{Rioul17} to prove Shannon's vector EPI. We find it also convenient to prove the complex matrix EPI:

\begin{theorem}
The matrix EPI~\eqref{mepi} holds  for any $m\times n$ complex matrix $\A$ and any random vector $X$ of independent complex components $X_i$, where 
$X^*$ is a c-normal vector with independent components $X^*_i$ satisfying~\eqref{sameentropies}. If equality holds in~\eqref{mepi} then
all unrecoverable components present in $\A X$ (in the sense of Definition~\ref{presrecov}) are normal.  
\end{theorem}
The exact necessary and sufficient condition for equality is more involved and will appear elsewhere.

\begin{IEEEproof}
We sketch the proof by going through the above proofs in Sections~\ref{proofsection} and~\ref{eqsection} and pointing out the differences:

\S\ref{A}: 
The scaling property of entropy now reads
$h(\A X)=h(\hat{\A}\hat{X})=h(\hat{X})+\log|\hat{\A}| = h(X)+\log|\A|^2$.

\S\ref{B}:
Since $h(X^*)=\log \bigl( \pi e \sigma^2 \bigr)$ for $X^*\sim\mathcal{CN}(0,\sigma^2)$, independent $X^*_j$ with equal entropies are i.i.d. 

\S\ref{C}: The Gram-Schmidt orthonormalization takes place in $\C^n$ with $h(\A' X)= h(\A X) + \log|\L|^2$.

\S\ref{D}: $\U=\bigl(\!
\begin{smallmatrix}
\A\hphantom{'}\\[0.5ex]\hline \A' 
\end{smallmatrix}
\!\bigr)$
is now an $n\times n$ \emph{unitary} matrix. 
Recall that a circularly symmetric $X^*\sim\mathcal{CN}(0,\K)$ is such that $\A X^*\sim\mathcal{CN}(0,\A\K\A^\dag)$ for any $\A$.
Since $X^*\sim\mathcal{CN}(0,\sigma^2\I)$ is i.i.d., $\U X^*\sim\mathcal{CN}(0,\sigma^2\U\U^\dag=\sigma^2\I)$ is also i.i.d. and the inverse transformation is the conjugate transpose $X^*=\A^\dag  \tilde{X} + {\A'\,}^\dag  \tilde{X}' $.

\S\ref{E}: Lemma~\ref{bm} replaces Lemma~\ref{nt} and~\eqref{jacobiandiag} becomes 
\begin{equation}
T'(\hat{X^*}) =  \diag \bigl(T'_1(\hat{X^*_1}), \ldots, T'_n(\hat{X^*_n})\bigr)
\end{equation}
in \emph{block-diagonal} form where each $2\times 2$ block $T'_i(\hat{X^*_i})>0$ is symmetric positive definite.


\S\ref{G}: In terms of the hat variables: 
\begin{equation}\label{Elog|T'|zero}
\E \log |T_j'(\hat{X^*_j})|=0\qquad (j=1,2,\ldots,n). 
\end{equation}
where $|\cdot|$ denotes the absolute value of the 
 determinant, 
 and
\begin{align}
h\bigl(\hat{\A} &T(\hat{\A}^t  \hat{\tilde{X}} + \hat{\A'\,}^t  \hat{\tilde{X}'} ) \mid  \hat{\tilde{X}'} \bigr)\notag\\
&= h(\tilde{X}) + \E \log | \hat{\A} T'(\hat{X^*}) \hat{\A}^t|
\end{align}

\S\ref{H}: We show that Lemma~\ref{logconcmatrixlemma} still holds when $\pmb{\Lambda}$ is  block-diagonal with  $2\times2$ diagonal blocks $\lambda_j>0$ (symmetric positive definite). Write
\begin{equation}
\lambda_j = \hat{u_j}\, d_j \, \hat{u_j}^t
\end{equation}
where 
$d_j$ is $2\times 2$ diagonal with positive diagonal elements and $\hat{u_j}$ is a rotation matrix, corresponding to a complex unit $u_j = e^{i\theta_j}$. Then the block-diagonal $\hat{\U}=\diag(\hat{u_1},\ldots,\hat{u_n})$ is orthonormal 
 and $\D=\diag(d_1,\ldots,d_n)$ is diagonal. We can now apply Lemma~\ref{logconcmatrixlemma} to $\hat{\A}\hat{\U}$ 
 and 
 $\D$: 
\begin{align}
\log|\hat{\A}\pmb{\Lambda}\hat{\A}^t | 
&\geq \tr (\hat{\A}\hat{\U} [\log\D] \hat{\U}^t\hat{\A}^t) \label{newineq}\\
&= \tr (\hat{\A} [\log \pmb{\Lambda}] \hat{\A}^t)
\end{align} 
where $\log \pmb{\Lambda}$ is the (block diagonal) logarithm of 
$\pmb{\Lambda}>0$. Thus
\begin{align}
\tr (\hat{\A} [\log \pmb{\Lambda}] \hat{\A}^t)&=\sum\nolimits_i\tr(\sum\nolimits_j \hat{A_{i,j}}  
[\log \lambda_j]
\hat{A_{i,j}}^t)\\
&={\sum\nolimits_{i}\sum\nolimits_{j}} |A_{i,j}|^2 \tr(\log \lambda_j) 
\end{align}
where $\tr(\log \lambda_j)=\log |\lambda_j|$ since $\lambda_j$ is symmetric positive definite.
Thus we obtain
\begin{equation}
\llap{\ensuremath\E} \log | \hat{\A} T'\!(X^*)\!\hat{\A}^t| \geq\!  {\sum\nolimits_i\!\sum\nolimits_j\!} |A_{i,j}|^2 \E\log |T'_j(\hat{X^*_j})|{=0}
\end{equation}
which is the final step to prove the (complex) matrix EPI~\eqref{mepi}.

Assume that equality holds in~\eqref{mepi} as in the converse part of the proof of Theorem~\ref{thm-eq} (Section~\ref{eqsection}). That proof is unchanged
up to the point where one considers the equality condition in Lemma~\ref{logconcmatrixlemma} applied to $\hat{\A}\hat{\U}$ and diagonal $\D$, that is, in~\eqref{newineq}. By the strict concavity of the logarithm, equality holds in~\eqref{newineq} if and only if for any two nonzero elements in the same row of $\hat{\A}\hat{\U}=\hat{\A\U}$, the corresponding two diagonal elements of $\D$ are equal. Since $\U=\diag(e^{i\theta_1},\ldots,e^{i\theta_n})$, the nonzero elements of $\A\U$ are at the same places as those of $\A$, where $\A$ is of the form~\eqref{canonical}. Therefore, due to the structure of $\hat{\A\U}$,
for any $j$ such that $r<j\leq n$, the two diagonal elements of $d_j$  are equal to the two diagonal elements of another $d_k$ where $r<k\leq n$, $k\ne j$,
which implies  $\lambda_j=\lambda_k$. 
This gives~\eqref{dereqae} from which one concludes as before that for all $r<j\leq n$, $T_j$ is linear, and, therefore, $X_j=T_j(X^*_j)$ is normal.
\end{IEEEproof}

\section{Application to Blind Source Extraction}

The theoretical setting of the blind source extraction problem is as follows~\cite{Vrins07}. We are given $n$ (zero-mean) independent (real or complex) ``sources'' $X=(X_1,X_2,\ldots,X_n)^t$ which are mixed using an $n\times n$ invertible (real or complex) matrix~$\M$, resulting in the observation $Y=\M X$. The covariance matrix $\K_Y$ of $Y$ can be estimated but both $\M$ and $X$ are unknown. Since one can introduce arbitrary scaling factors in $\M$ and $X$ for the same observation~$Y$, we can assume an arbitrary normalization of the sources. For convenience we assume here that they have the same entropies:
\begin{equation}\label{hequal}
h(X_1)= h(X_2)=\cdots=h(X_n).
\end{equation}
Blind source extraction (or partial BSS) of $m$ sources ($1\leq m\leq n$) aims at finding a (full rank) $m\times n$ matrix $\W$ such that $Z=\W Y$ is composed of $m$ (out of $n$) original sources, up to order and scaling. In other words $\A=\W\M$ should have exactly one nonzero element per row.

\begin{definition}[Contrast function~\cite{Vrins07}]
A \emph{contrast} $\mathcal{C}(\W)$ is a function that is invariant to permutation and scaling of the rows  $\w_i$ of $\W$, and such that it achieves a \emph{minimum} if only if $\A=\W\M$ has one nonzero element per row. 
\end{definition}

\begin{theorem}
Assume that \emph{at most one} source is normal. 
Then\smallskip
\begin{equation}
\mathcal{C}(\W) =
\smash{\sum_{i=1}^m} h(\w_iY) -  \frac{1}{2} \log |\W\K_Y \W^t|  
\end{equation}
where $\w_i$ are the rows of $\W$, is a contrast function.
\end{theorem}
Such a contrast function was first proposed by Pham~\cite{Pham06} (see also~\cite{CrucesCichockiAmari01}) in the real case with a different proof that uses the classical EPI  for $m=1$ and Hadamard's inequality. 
It is particularly interesting to rewrite it in terms of the matrix EPI:
\begin{IEEEproof}
The real and complex cases being similar, we prove the result in the real case.
Let $\A=\W\M$ and let $X^*$ be as in~\eqref{mepi}. For i.i.d. components we can rewrite~\cite[Eq.~(13)]{ZamirFeder93} as $h(\A X^*)= mh +\frac{1}{2}\log|\A \A^t|$ where $h$ is the common value of~\eqref{hequal}. Since $Z=\W Y=\A X$, up to an additive constant we may decompose $\mathcal{C}$ as
\begin{equation}\label{cich}
\mathcal{C}(\W) = \mathcal{C}_h(\W)+\mathcal{C}_i(\W)  + \mathrm{Cst.}
\vspace*{-1ex}\end{equation}
where\vspace*{-1ex}
\begin{align}
\mathcal{C}_h(\W)&= h(\A X)  - h(\A X^*) \geq 0\\
\mathcal{C}_i(\W)&= \sum_i h(Z_i) \;-\; h(Z) \geq 0
\end{align}

The term $\mathcal{C}_i(\W)$ is minimum (with minimum value $=0$) if and only if the components $Z_i$ of $Z$ are independent.

The $\mathcal{C}_h(\W)$  is minimum (with minimum value $=0$)  if and only if equality holds in~\eqref{mepi}. Since at most one source is normal,  at most one source present in $\A X$ can be unrecoverable. But if one (normal) source is not recoverable, the canonical form~\eqref{canonical} implies that at most one column of $\A_u$ is nonzero, which contradicts the maximality of $r$ in Lemma~\ref{lem-can}. Therefore, $r=m$ and the canonical form of $\A$ becomes $(\I_m\mid 0)$.

With the additional constraint $\mathcal{C}_i(\W)=0$ that components of $Z=\A X$ are independent, it follows from the Darmois–Skitovich theorem~\cite{Comon94} (see~\cite{ErikssonKoivunen06} in the complex case) that $\A$ has exactly one nonzero per row. 
\end{IEEEproof}
Interestingly, the contrast function in the form~\eqref{cich} represents a transition between the two well-known extreme cases:
\begin{itemize}
\item
$m=1$, for which $\mathcal{C}_i=0$ where each source is extracted one by one using the classical EPI (minimize $\mathcal{C}_h$); 
\item
$m=n$, for which $\mathcal{C}_h=0$, where all $n$ sources are separated simultaneously; we are then reduced to an independent component analysis (ICA) problem~\cite{Comon94,ErikssonKoivunen06} in which the multivariate ``mutual information''~$\mathcal{C}_i=D(p(Z)\|\prod_i p(Z_i))$ is minimized.
\end{itemize}




%



\bibliographystyle{IEEEtran}
\bibliography{MEPI}\vspace*{0pt}

%
%

\end{document}